\documentstyle[PASJadd]{PASJ95}

\begin{document}
\title{
CO Observations of Luminous IR Galaxies at Intermediate Redshift 
}
\author{Yoshinori {\sc Tutui}\thanks
{Present address: {\it NHK}~(Japan Broadcasting Corporation), Program
Production Department, Science Programs Division, 
2-2-1 Jinnan, Shibuya-ku, Tokyo 150-8001, Japan},$^1$ 
Yoshiaki {\sc Sofue},$^1$
Mareki {\sc Honma},$^{2,3}$ Takashi {\sc Ichikawa},$^{4}$
and Ken-ichi {\sc Wakamatsu}$^5$
\\ [12pt]
$^1$ {\it Institute of Astronomy, University of Tokyo, 
	Mitaka, Tokyo 181-8588, Japan}\\
{\it E-mail (YT): tutui@mtk.ioa.s.u-tokyo.ac.jp}\\
$^2$ {\it VERA Project Office, National Astronomical Observatory of Japan
	Mitaka, Tokyo 181-8588, Japan}\\
$^3$ {\it Mizusawa Astrogeodynamics Observatory, 
        National Astronomical Observatory of Japan, 
        Mizusawa, Iwate 023-0861, Japan}\\ 
$^4$ {\it Astronomical Institute, Tohoku University, 
	Aoba, Sendai 980-8578, Japan}\\
$^5$ {\it Department of Technology, Gifu University, 
	1-1 Yanagido, Gifu 501-11, Japan}
}
\abst{
We present new measurement of $^{12}$CO($J=1-0$) emission
from 16 luminous infrared galaxies (LIGs) at intermediate 
redshift ($cz \sim 10,000 - 50,000 {\rm km~s^{-1}}$).
These new data were selected by isolated and normal morphology.
Although there already exist measurements of CO 
emission from LIGs in the literature,
they are mostly strongly interacting/merging system.
The new CO data represent an important new addition
to the literature in that they both expand the relatively 
small sample of LIGs measured in CO,
and they include the interesting subset of of LIGs  
that were selected by isolated and normal morphology.
The CO observations were performed
using the NRO 45-m telescope.
From the measurement of CO emission and the IRAS database, 
we discuss the molecular gas and dust properties
of late-type galaxies at intermediate redshift.
Comparison of the CO and dust properties 
of the new result with those from other CO measurements
revealed characteristics of this sample:
(1) It is the deepest CO observations of IRAS galaxies
at intermediate redshift without strong interaction features.
(2) It has typical properties of normal IRAS galaxies 
in terms of star-formation efficiency, color-color diagrams
and galactic nuclear activity. 
(3) It has smaller gas-to-dust ratio than normal IRAS galaxies.
This can be explained by two-component dust model,
and our sample consists of most of warm dust.
}
\kword{Galaxies: spiral --- Galaxies: distances and redshifts 
--- ISM: molecules}
\maketitle
\thispagestyle{headings}

\section{Introduction \label{sect_introduction_IRASgal}}

The Infrared Astronomical Satellite (IRAS ) 
revealed a number of extragalactic objects
with luminosity dominated by FIR emission (e.g. Soifer et al. 1984).
IRAS  surveyed 96\% of the sky, with a completeness limit of 
$\sim 0.5$ Jy at 12 $\mu\rm m$, 25 $\mu\rm m$ and 60 $\mu\rm m$,
and  $\sim$ 1.5 Jy at 100 $\mu\rm m$,
and with angular resolution $\sim$ 0.5 $\hbox{$^\prime$}$ for 12 $\mu\rm m$
$\sim$ 2 $\hbox{$^\prime$}$ for 100 $\mu\rm m$.
It discovered $\sim$ 20,000 galaxies, which 
had not been previously cataloged. 
The majority of IRAS  extragalactic galaxies
are late-type galaxies.
Galaxies detected by IRAS , whose luminosity is dominated
by FIR emission, are called IRAS  galaxies,
and are categorized 
as luminous infrared galaxies (LIGs) for infrared luminosity
of $L_{\rm IR} > 10^{11} {L}_{\odot}$,
ultraluminous infrared galaxies (ULIGs) for 
$L_{\rm IR} > 10^{12} {L}_{\odot}$,
and hyperluminous infrared galaxies (HyLIGs) for
$L_{\rm IR} > 10^{13} {L}_{\odot}$.
Infrared luminous galaxies whose luminosity 
is more than $10^{11} {L}_{\odot}$ 
become the dominant population at the intermediate redshift 
of  $z \ltsim 0.3$ (Sanders \& Mirabel 1996).
The fraction of merger and interacting systems among IRAS  galaxies 
increases with infrared luminosity;
that is $\sim 12\%$ for $\log(L_{\rm IR}/{L}_{\odot}) = 10.5 - 11$,
$\sim 32\%$ for $\log(L_{\rm IR}/{L}_{\odot}) = 11 - 11.5$,
$\sim 66\%$ for $\log(L_{\rm IR}/{L}_{\odot}) = 11.5 - 12$ and 
$\sim 95\%$ for $\log(L_{\rm IR}/{L}_{\odot}) > 12$  
(Sanders \& Mirabel 1996).

CO-line observations have revealed that LIGs 
are extremely rich in molecular gas.
Early CO observations of infrared selected galaxies
with $L_{\rm IR} = 10^{10} - 10^{11} {L}_{\odot}$ showed 
a rough correlation between CO and FIR luminosity
(Young et al. 1984; Young et al. 1986).
A number of single-dish CO surveys of IRAS  galaxies 
have been performed (Sanders et al. 1991, Mirabel et al. 1990, 
Tinney et al. 1990, Downes et al. 1993, Mazzarella et al. 1993,   
Young et al. 1995, Elfhag et al. 1996, Solomon et al. 1997),
which consist of galaxies 
with IR luminosities of $L_{\rm IR} = 10^{10} - 10^{13} {L}_{\odot}$ 
from nearby to intermediate redshift.
Correlations between CO and IR luminosities
are also found for these samples.
Furthermore, it is found that the IR-to-CO luminosity ratio
($L_{\rm IR}/{L}_{\small{\rm CO}}'$), which represents the star formation efficiency
in a galaxy, increases with the CO luminosity,
suggesting that molecular-gas-rich galaxies show
a high star-forming efficiency.
Assuming the Galactic $^{12}$CO($J=1-0$)-H2 conversion factor of
$N_{\rm H_{2}}/{I}_{\small{\rm CO}} = 3.0 \times 10^20  {\rm cm^{-2} (K km s^{-1})^{-1}}$,
the total molecular-gas mass for ULIGs is as high as 
$M_{\rm H_{2}} \gtsim 10^{10} M_{\odot}$,
and the star-formation efficiency in ULIGs is,
on average, much higher than 
any of the most active star-forming Galactic GMC cores.

CO observations of ULIGs
at intermediate redshift ($cz \sim 10,000 - 50,000 {\rm km~s^{-1}} $) 
have been performed by
Mirabel et al.(1990) using the SEST 15-m telescope, 
Sanders et al.(1991) using the NRAO 12-m telescope,
Solomon et al.(1997) using the IRAM 30-m telescope and 
Lavezzi \& Dickey (1998) using the NRAO 12-m telescope
(see Sanders \& Mirabel 1996, for more references).
In addition to the above, we have performed CO observations
using the NRO 45-m telescope.
CO observations for ULIGs at intermediate redshift are important
for the following points:
(1) Objects at intermediate redshift 
($z \sim 0.1 - 1$) are significant for galactic evolution
from the most-active epoch at $z \sim 1 -2$,
to the local Universe at $z \sim 0$.
(2) Galaxies at intermediate redshift are the most distant targets
in which we can detect the CO-line from the galactic disk,
which is related to the global properties 
of the molecular gas in a galaxy.
(3) High ratio of merger or interacting systems 
and high star-formation efficiency of ULIGs 
can help reveal the star-formation history.
(4) Strong CO emission and stable CO linewidths
can be used for the Tully-Fisher relation
to measure distances to galaxies at intermediate redshift.
Photometric observations for the CO-line Tully-Fisher relation 
have been performed using the Okayama Astrophysical Observatory 1.88-m
telescope, and the results will be discussed in a forthcoming paper.

Results of our CO-line observations 
and definitions of CO and FIR properties discussed in this paper
are described in section 2.
We compare our sample with these CO observations to evaluate
the characteristics of our sample in section 3.
The results are discussed in section 4, 
and summarized in section 5.
We also discuss the samples in terms of the IRAS  color-color diagrams
and galactic activities in appendices 1 and 2.
Throughout this paper we use $H_0 = 75 ~{\rm km ~s^{-1} ~Mpc^{-1}}$
and $q_0 = 0.5$.

\section{Data and CO observations \label{sect_obs_NRO45m}}

$^{12}$CO($J=1-0$) ~line observations of galaxies at intermediate redshift were 
obtained using the Nobeyama Radio Observatory (NRO) 45-m telescope.
We also use the data from the literature (Solomon et al. 1997,
Sanders et al. 1991, Mirabel et al. 1990 and Lavezzi \& Dickey 1998).
In comparing these samples, 
we used given integrated intensities 
and the same formulae to estimate CO luminosities, or molecular gas
mass.

\subsection{Sample selection}

The main purpose of our CO observations was to obtain 
CO linewidths in order to apply them
to the CO-line Tully-Fisher relation 
to measure distances to galaxies 
and determine the Hubble constant at intermediate redshift.
The target was late-type galaxies 
with bright CO-line emission and 
isolated normal morphology located at intermediate redshift.
The sample selection was made using the following criteria.

(1){\it Redshift criterion.}
Redshift criterion of the sample was
$cz = 10,000 - 20,000 {\rm km~s^{-1}} $ in 1994/1995 observations,
and $ cz = 20,000 - 30,000 {\rm km~s^{-1}} $ in 1995/1996 observations.
After we confirmed the possibility of CO detection
at such redshift, 
we set the redshift criterion to $cz = 30,000 - 50,000 {\rm km~s^{-1}} $
in 1996/1997 observations.

(2){\it FIR flux density criterion.}
In order to obtain sufficient CO emission at intermediate redshift
we selected relatively strong FIR-emission sources 
at 60 $\mu\rm m$ and 100 $\mu\rm m$,
because FIR emission, which is dominated by thermal emission
of dust heated by surrounding starlight in UV,
is related to CO emission from dense molecular-gas regions.
We selected galaxies whose IRAS  flux densities at 
60 $\mu\rm m$ and 100 $\mu\rm m$ are greater than 1 Jy.

(3){\it Morphology criterion.}
Although morphology information at intermediate redshift
was not sufficient in catalogs, 
we judged the morphology of the sample 
using images from the STScI Digitized Sky Survey (DSS).
Strong interacting galaxies and mergers 
were excluded from the sample.
However, weakly interacting galaxies,
whose tails or irregular arms were not resolved by the DSS images
at the resolution limit of $\sim 1\hbox{$^{\prime\prime}$}$,
were included,
because influences of weak interaction do not affect the 
CO linewidth so much as for the HI linewidth (Tutui \& Sofue 1997).

(4){\it Position error criterion.}
Since the half-power beam width (HPBW) 
of the NRO 45-m telescope was 15\hbox{$^{\prime\prime}$} at the frequency of 
$^{12}$CO($J=1-0$), 115.271204 GHz in the object rest frame, 
galaxies whose position error  
listed in the NASA Extragalactic Database(NED)
is less than 10\hbox{$^{\prime\prime}$} were selected. 
We also cross-checked the position using the DSS images.

(5){\it Recession velocity error criterion.}
The band width of the $^{12}$CO($J=1-0$) ~observation of the NRO 45-m telescope
was 250 MHz, corresponding to $650(1+z)~ {\rm km~s^{-1}} $. 
Since linewidths of observed galaxies were expected to be about 
$200 - 500 {\rm km~s^{-1}} $, the recession velocity error criterion
to select galaxies was less than 100 ${\rm km~s^{-1}} $.
The recession velocity was taken from
the IRAS  redshift surveys by  
Strauss et al. (1992) and Fisher et al. (1995).

\subsection{CO observations}

CO-line observations were performed as a NRO 45-m long-term project,
and were carried out on January 14 to 23 and December 9 to 12 in 1994, 
January 6 to 10, March 13 to 17 and December 17, 18, 21, 22 in 1995,
January 17 to 22, February 18 to 22 and December 2 to 5 in 1996,  
and January 8 to 12 in 1997.
The HPBW of the NRO 45-m telescope was
15\hbox{$^{\prime\prime}$} at the frequency of $^{12}$CO($J=1-0$) ~line,
and the aperture and main-beam efficiencies were $\eta_{\rm a}$
= 0.35 and $\eta_{\rm mb}$ = 0.50, respectively,
as measured by observing the planets Mars and Saturn.
As the receiver frontends, we used cooled SIS 
(superconductor-insulator-superconductor)
receivers, 
which could receive two orthogonal polarizations simultaneously,
with a SSB filter to select one of the sidebands.
The receiver backends were 2048-channel wide-band 
acousto-optical spectrometers (AOS).
The total channel number corresponds to frequency widths of 
250 MHz, and therefore, to a velocity coverage in the rest frame
of the galaxy of $\sim 650 ~{\rm km~s^{-1}}$.
%
%
The center frequency was set to the 1024-channel, which 
corresponded to $115.271204~(1+z)^{-1}$ for each galaxy.
The system noise temperature was 300 $-$ 800 K in the single side band
at the observing frequencies.
Calibration of the line intensity was made using an absorbing 
chopper in front of the receiver, yielding an antenna temperature
($T_{\rm A}^{\ast}$), corrected for both the atmospheric and antenna ohmic losses.
The intensity scale of $T_{\rm A}^{\ast}$ was converted to the main-beam 
brightness temperature by $T_{mb}=T_{\rm A}^{\ast}/ \eta_{\rm mb}$.
Subtraction of  sky emission was performed 
by on-off position switching, and the offset of the off-position
was 5$\hbox{$^\prime$}$  away from the on-position.
On-source total integration time for each galaxy 
ranged from 30 minutes to 3 hours.
Antenna pointing was performed by observing nearby SiO maser
sources at 43 GHz every 60 $-$ 90 minutes.
The pointing accuracy was better than $\pm 4\hbox{$^{\prime\prime}$}$ during all 
observations.
Total observation time for the on/off position integrations 
and pointing was about 90 minutes to 9 hours for individual galaxies.
After flagging bad spectra, subtraction of the baseline
was performed by linear-baseline fitting.  
Adjacent channels were binned to a velocity resolution of 10 ${\rm km~s^{-1}} $.
Noise level of the resultant spectra at velocity resolution of 10
${\rm km~s^{-1}} $ was 2 $-$ 5 mK in $T_{\rm A}^{\ast}$.

\subsection{CO-line profiles}

We have obtained CO-line spectra of sixteen galaxies 
at intermediate redshifts, $cz \sim 10,000 - 50,000 {\rm km~s^{-1}} $,
which are listed in table 1.
These galaxies are some of the most distant examples of 
IRAS  galaxies whose CO-line is detected.
The observed CO-line profiles are shown in figure 1.
Measurement of CO integrated intensities and linewidths
were performed carefully,
not only by using the final CO-line
profiles but also by comparing CO-line profiles 
between individual spectrometers and each run.
In table 2 we list the observed 
properties of the CO-line profiles; 
on-source integration time, the r.m.s. antenna temperature,
CO linewidths at the observed and rest frames, 
the antenna temperature at the peak level,
and the integrated intensity corrected for the main beam efficiency.




\subsection{CO luminosities and molecular gas properties 
\label{sect_COproperty}}

CO luminosity is expressed mainly in two ways.
One is formulated by an integrated flux density with the unit 
of ${L}_{\odot}$.
The other is formulated by integrated intensity, or sometimes an 
integrated flux density, with the unit of ${\rm K km s{}^{-1} pc{}^{2}}$.
Observed integrated intensity is described by

\begin{equation}
I_{\nu_{\rm obs}} = \int T_{mb} d{V}_{\rm obs},
\end{equation}

\noindent
where $T_{mb}$ is the main beam temperature related to the 
antenna temperature $T_{\rm A}^{\ast}$ and the main beam efficiency $\eta_{mb}$,
as $T_{mb} = T_{a}^{\ast}/\eta_{mb}$.
We should note the integrated intensity for a point source object
in comparing between different measurements,
because the intensity depends on the beamsize and efficiencies. 
For example, the observed integrated intensity of PG0157 (=Mrk1014),
which should essentially be a point source for both 
the NRO 45-m and IRAM 30-m telescopes,
are 5.11 ${\rm K~km~s^{-1}}$ for NRO 45-m and 1.8 ${\rm K~km~s^{-1}}$ 
for IRAM 30-m in Solomon et al.(1997),
but it is consistent within 50\% of the integrated intensity
for the efficiency of $\eta_{\rm a} = 0.35$ and $\eta_{\rm mb} = 0.5$
for NRO 45-m, and $\eta_{\rm a} = 0.3$ and  $\eta_{\rm mb} = 0.6$
for IRAM 30-m.

CO luminosity ${L}_{\small{\rm CO}}'$ in the unit of ${\rm K~km~s^{-1}~pc^{2}}$
is denoted with the observed integrated intensity $I_{\nu_{\rm obs}}$
and the beam solid angle $\Omega_{b}$ as

\begin{equation}
{L}_{\small{\rm CO}}' = \frac{\Omega_{b} I_{\nu_{\rm obs}} {D}_{{\small L}}^{2}}{(1+z)^{3}}, 
\label{obs45m.eq.Lco'}
\end{equation}

\noindent
where $z$ is the redshift and 
${D}_{{\small L}}$ is the luminosity distance formulated 
using the Hubble constant
$H_0$ and the deceleration parameter $q_{\rm 0}$ as

\begin{equation}
{D}_{{\small L}} = \frac{c}{H_0 q_{\rm 0}^{2}} 
\left ( q_{\rm 0} z + (q_{\rm 0} - 1)( \sqrt{2 q_{\rm 0} z + 1} -1 ) \right ).
\end{equation}

\noindent
Mass of molecular gas is estimated from CO emission 
assuming the Galactic CO-to-H$_{2}$ conversion factor.
We adopt a conversion factor of 
$N_{\rm H_{2}} / I_{CO} =  
3.0 \times 10^20  {\rm cm^{-2} (K km s^{-1})^{-1}}$
(Strong et al. 1988; Solomon et al. 1991).
The correction factor from  $H_{2}$ mass 
to molecular gas mass including He and other elements 
is referred from Allen (1973), as

\begin{equation}
M_{\rm gas} = 1.36 \times M_{\rm H_{2}}.
\end{equation}

\noindent
We estimated CO properties of the galaxies 
with the results of the CO observations 
listed in table 3.
Ranges of the CO luminosity and molecular gas mass
of our sample are
${L}_{\small{\rm CO}}' = 5.5 \times 10^{8} - 7.4 \times 10^{9}$ 
and $M_{\rm gas} = 3.3 \times 10^{9} - 4.6 \times 10^{10}$,
respectively.
These are about 20 times greater than the CO luminosity and 
molecular gas mass of the Milky Way interior to the solar circle
(Solomon \& Rivolo 1989).

\subsection{FIR luminosities and dust properties}

FIR luminosities of LIGs are normally
adopted by the formula using the IRAS  60 and 100 $\mu\rm m$ 
flux densities, 

\begin{equation}
\left ( \frac{L_{\rm FIR}}{{L}_{\odot}} \right) = 3.92 \times 10^{5} 
\left \{
  2.58 \left ( \frac{S_{60}}{\rm Jy} \right )
+      \left ( \frac{S_{100}}{\rm Jy} \right )
\right \} 
C \left ( \frac{{D}_{{\small L}}}{\rm Mpc} \right )^{2},
\label{eq_Lfir}
\end{equation}

\noindent
where the constant $C$ is a correction factor 
of the flux density beyond the range of $60-100 \mu\rm m$,
which is typically in the range of 1.5 $-$ 2.1
(Solomon et al. 1997), and here we assumed $C  = 1.8$.

Dust mass is estimated by the dust temperature 
$T_{d}$ derived from flux densities
at 60 $\mu\rm m$ and 100 $\mu\rm m$ of the blackbody radiation.
Assuming that the dust emission follows the emissivity law
of $S_{\nu} \propto \lambda^{-1}$,
the dust mass $M_{\rm dust}$ is approximately expressed as (Hildebrand 1983), 

\begin{equation}
M_{\rm dust} = 4.589 \left (\frac{S_{100}}{{\rm Jy}} \right ) \left
(\frac{{D}_{{\small L}}}{{\rm Mpc}} \right )^{2} \left ( \exp(144/T_{d}) -1 \right)
\label{eq_Md}
\end{equation}

\noindent
where, $T_{d}$ is given by the relation,
\begin{equation} 
\frac{S_{60}}{S_{100}} = 7.72 \frac{\exp(144/T_{d})-1}{\exp(240/T_{d})-1}.
\end{equation}

\noindent
Ranges of FIR luminosity and dust mass 
of our sample are $L_{\rm FIR} = 5.5 \times 10^{10} - 2.57 \times 10^{12}$ 
and $M_{\rm dust} = 1.7 \times 10^7 - 1.1 \times 10^8 M_{\odot}$, respectively.
The dust properties of our sample are listed in table 4.



\section{Comparison with other CO-line samples
\label{sect_result_IRASgal}}

\subsection{Individual CO-line samples}

We compared our results (hereafter NRO sample)
with other CO-line observations
which had been obtained for a number of IRAS  galaxies
at intermediate redshift.
We used the following four CO samples data for the comparison,
and hereafter the samples are denoted as
IRAM sample (Solomon et al. 1997), 
NRAO-S sample (Sanders et al. 1991), 
SEST sample (Mirabel et al. 1990) 
NRAO-L sample (Lavezzi \& Dickey 1998).
Sampling of IRAS  galaxies at intermediate redshift
is based on the FIR flux density,
because FIR flux density is generally correlated with 
CO flux density.
Figures 2 and 3 show
plots of redshift $cz$ against FIR luminosity 
and against CO luminosity for all the samples, respectively.
The majority of galaxies in the IRAM sample are 
ULIGs with higher CO luminosity at farther intermediate redshift,
and are interacting galaxies or mergers.
NRAO-S sample and SEST sample 
made pioneering surveys of LIGs 
in the northern and southern skies,
respectively, and they were relatively shallower samples.
NRAO-L sample was a sample of IRAS  galaxies 
in clusters of galaxies 
at relatively lower $cz$. 
Our NRO sample shows the deepest CO observations
at intermediate redshift.
The purpose of our observations was obtaining 
CO linewidths of isolated galaxies 
with normal morphology at intermediate redshift
in order to apply them to the CO Tully-Fisher relation.
Although the CO emission of normal galaxies is generally 
not as bright as for ULIGs,
deep observations with long integration 
have obtained CO emission from galaxies at intermediate redshift.
Since the purposes of individual  CO observations were different, 
the properties of target galaxies in each sample 
also show various features.
Remarks and details of individual samples are as follows
and are also listed in table 5.




\subsection{CO and FIR luminosities}

A correlation between CO and FIR luminosities 
for LIGs was found 
by airborne FIR observations (Rickard \& Harvey 1984),
and by IRAS  observations (Young et al. 1984; Sanders \& Mirabel 1985).
Figure 4 is a plot of CO luminosity
against FIR luminosity for all the samples,
showing the correlation between CO and FIR luminosities.
The ratio of $L_{\rm FIR}/{L}_{\small{\rm CO}}'$, or $L_{\rm FIR}/M_{\rm H_{2}}$,
represents the star formation efficiency (SFE).
The correlation shows that SFE for LIGs  
is higher than that for nearby normal spiral galaxies.
Nearby samples presented by Sage (1993) 
and Solomon \& Sage (1988) are indicated as a reference 
by the solid thick line and the solid thin line, respectively. 
The former is a distance-limited sample,
whereas the latter is a flux-limited sample, 
and the flux-limited sample shows a larger SFE.
$L_{\rm FIR}/{L}_{\small{\rm CO}}'$ for giant molecular clouds (GMCs) in the Milky Way
is distributed between the two dotted lines 
(Solomon et al. 1986; Mooney \& Solomon 1988).
Most of the galaxies in the samples have larger values of 
$L_{\rm FIR}/{L}_{\small{\rm CO}}'$ than that of the GMCs.
As Solomon \& Sage (1988) and Solomon et al. (1992) indicated,
$L_{\rm FIR}/{L}_{\small{\rm CO}}'$ for LIGs is not constant 
but increases with the CO luminosity, as seen in
figure 4.
We fitted all the samples by a power law and obtained 
a relation between FIR and CO
luminosities given by,

\begin{equation}
\log L_{\rm FIR} = 1.41 \log {L}_{\small{\rm CO}}' - 1.60 ~~(\rm{all~samples}).
\end{equation}

\noindent
Thus, all the samples which we examined in this paper
have systematically larger SFE than the nearby normal spirals.
The IRAM sample  shows a remarkably higher SFE,
whereas NRAO-L sample  is not far from that of GMCs.
Other samples are distributed between them uniformly.
The plots of CO against FIR luminosities for the individual samples
are shown in figure 5.



\subsection{Gas and Dust Properties}

As indicated in equation (7),
the dust mass in galaxies is generally estimated using 
the IRAS  flux at 60 and 100 $\mu\rm m$,
adopting a single-temperature dust model.
However, this estimation overlooks cold dust ($10-20$K) radiation, 
because the radiation of cold dust is dominant at $\lambda > 100 \mu\rm m$,
and thus the gas-to-dust ratio is overestimated.
The gas-to-dust ratio derived from the IRAS  flux 
for nearby spiral galaxies  
is typically $\sim$ 570 (Young et al. 1986; Young et al. 1989; 
Stark et al. 1986).
On the other hand, the Galactic local gas-to-dust ratio is
derived from the gas column density and color excess of nearby stars,
and therefore should be close to the real gas-to-dust ratio of 
$100-150$ (e.g. Spitzer 1978; Bohlin et al. 1978).
It has been suggested that the discrepancy of the gas-to-dust ratio
is explained by two dust components, namely warm and cold dust.
The two-component dust model, which consists of 
cold ($10-20$K) dust associated with quiescent molecular clouds
and warm ($30-60$K) dust associated with star-forming regions,
is considered (e.g. de Jong \& Brink 1987).
For nearby spiral galaxies
the discrepancy can be explained  when the dust consists of
$10-20\%$ of warm dust and $80-90\%$ of cold dust (Devereux \& Young 1990).
Figure 6 is a plot of gas mass against dust mass.
Although the NRAO-S, SEST and NRAO-L samples follow 
the gas-to-dust ratio of nearby galaxies ($\sim 570$),
the IRAM and NRO samples are distributed between values 
of nearby galaxies and the Galaxy.
It indicates that the warm dust is dominant 
for IRAM and NRO samples.


\subsection{Warm dust fraction \label{sect_w.dust.fract}}

Young et al. (1986) showed that FIR luminosity of a galaxy
depends on both CO luminosity and dust temperature.
As seen in figure 7, $L_{\rm FIR}/{L}_{\small{\rm CO}}'$ increases
with dust temperature.  
The IRAM sample  and NRO sample  have 
larger values of $L_{\rm FIR}/{L}_{\small{\rm CO}}'$ at any dust temperature
compared to the other samples.
We fitted the data in figure 7
by a power law 

\begin{equation}
L_{\rm FIR}/{L}_{\small{\rm CO}}' \propto {T_{d}}^{4.42}~~({\rm all~samples}), 
\label{eq_LfirLco--Td_all}
\end{equation}

\begin{equation}
L_{\rm FIR}/{L}_{\small{\rm CO}}' \propto {T_{d}}^{4.54}~~({\rm without IRAM and NRO samples}).
\label{eq_LfirLco--Td_without}
\end{equation}

\noindent
As discussed in the previous section, dust in galaxies 
consists of two dust components.
However, here we assume simply one dust component whose radiation 
is dominant in the FIR.
Then the FIR luminosity is described by

\begin{equation}
L_{\rm FIR} \propto M_{\rm dust} T_{d}^{4+n},
\end{equation}

\noindent 
where $n$ is the index of the dust emissivity law of $\lambda^{-n}$.
CO luminosity ${L}_{\small{\rm CO}}'$ is proportional to gas mass $M_{\rm gas}$,
and $L_{\rm FIR}/{L}_{\small{\rm CO}}'$ is described as

\begin{equation}
L_{\rm FIR}/{L}_{\small{\rm CO}}' \propto \frac{1}{f} T_{d}^{4+n},
\end{equation}

\noindent
where $f$ is the gas-to-dust ratio.
The power of the emissivity law is not well established 
and usually is adopted as $n=1$. 
Therefore, in estimating dust mass
and comparing the results with previous work,
we used the emissivity law of $n=1$.
In figure 7  the values of $L_{\rm FIR}/{L}_{\small{\rm CO}}'$ 
for galaxies in the IRAM and NRO samples are high.
It is recognized as due to the fact that 
the gas-to-dust ratio for these samples
is smaller than for the other samples.
This is consistent with 
dominant warm dust in the two-component dust model
discussed in the previous section.


\section{Discussion}

\subsection{CO beam coverage}
Smaller gas-to-dust ratios for the NRO and IRAM samples
may be induced by missing CO flux due to smaller beam size.
CO observations were performed by the 
single dish telescopes and pointed at the centers 
of target galaxies.
Assuming a cosmological model 
of $h_0 = 0.75$ and $q_0 = 0.5$,
the linear scale size o the $^{12}$CO($J=1-0$) ~beam (HPBW)
of the telescope along the redshift 
against angular size is denoted by,
\begin{equation}
D ({\rm kpc}) = 4.0 \times 10^2 ~\theta ({\rm arcsec})~ 
\frac{{(1+z)}^{1/2}-1}{{(1+z)}^{3/2}}. 
\end{equation}
The beam sizes of the samples are
HPBW $= 55\hbox{$^{\prime\prime}$}$ for NRAO 12-m telescope,
$44\hbox{$^{\prime\prime}$}$ for SEST 15-m telescope,
$30\hbox{$^{\prime\prime}$}$ for IRAM 30-m telescope,
and $15\hbox{$^{\prime\prime}$}$ for NRO 45-m telescope.
Even for NRO sample, which has the smallest beam,
the linear scale size at $cz > 10,000 {\rm km~s^{-1}} $ 
is greater than $\sim 10$ kpc,
which should cover the whole extent of CO-emitting disk of the galaxies.

\subsection{IRAS  $K$-correction} 
In estimating dust properties
we used observed IRAS  flux densities without $K$-correction. 
We here discuss the effects of $K$-correction.
Assuming a thermal spectral energy distribution 
and the emissivity law of $S_{\nu} \propto \lambda^{-1}$,
we estimate the effect of  $K$-correction on the IRAS  flux ratio 
at 60$\mu\rm m$ to 100$\mu\rm m$ and on the dust mass.
Dust temperature, dust mass and gas-to-dust ratio
are derived from flux densities at 60$\mu\rm m$ and 100$\mu\rm m$. 
Figure 8 shows a residual of 
the IRAS  flux ratio of 60$\mu\rm m$ to 100$\mu\rm m$ 
to which a $K$-correction is applied ($S~'_{60}/S~'_{100}$)
{\it to} the observed flux ratio without $K$-correction 
($S_{60}/S_{100}$).
Here flux densities to which $K$-correction is applied are 
denoted by  $S~'_{60}$ and $S~'_{100}$ 
The residual for galaxies 
at redshift $cz$ of 30,000 ${\rm km~s^{-1}} $ 
with dust temperature of 40 K 
is estimated to be $-0.1$.
Adopting this result, the effect of $K$-correction 
is found to be not significant
for the discussion of galaxy types in figure 10.
Figure 9 shows the dust mass ratio 
to which $K$-correction is applied {\it to} 
the observed dust mass without $K$-correction.
The effect of $K$-correction for most galaxies used in this study
is within 10 \% of the ratio.



\section{Summary \label{sect_summary_IRASgal}}

1.~~
We have performed long-integration $^{12}$CO($J=1-0$) ~observations 
for late-type galaxies
at intermediate redshift $cz \sim 10,000 - 50,000 {\rm km~s^{-1}} $
using the NRO 45-m telescope, and have obtained CO line profiles 
from 16 galaxies with isolated and normal aspects.
We compared the observed NRO sample  with other CO observations
of IRAS  galaxies at intermediate redshift,
[{\it i.e.} Solomon et al. 1997(IRAM sample);
Sanders et al. 1991(NRAO-S sample);
Mirabel et al. 1990(SEST sample);
Lavezzi \& Dickey 1998(NRAO-L sample)],
and found that the NRO sample  represents the deepest CO observations
at intermediate redshift.

2.~~
We compared the samples in terms of molecular gas and dust masses.
The gas-to-dust ratio for NRAO-S, SEST and NRAO-L samples are
equivalent to the values found in nearby spirals. 
On the other hand, the gas-to-dust ratio 
for the IRAM and NRO samples is smaller than for the other
samples and is close to the local gas-to-dust ratio of the 
Galaxy ($\sim 100-150$),
which is derived from gas column density and color excess of
nearby stars, 
and thus it is close to real gas-to-dust ratio.
The method to estimate dust mass using IRAS  flux at 
100 $\mu\rm m$ overlooks cold dust, 
and the gas-to-dust ratio tends to be overestimated.
For nearby spiral galaxies the gas-to-dust ratio is
typically 570, and it suggests that the dust components
consist of majority of cold dust and minority of warm dust.
This discrepancy can be explained if the dust component of the 
IRAM and NRO samples is mostly warm dust.

3.~~
We compared  the samples
in terms of IRAS  color-color diagrams and 
galactic nuclear activity, 
and found that there is no clear peculiarity 
for the NRO sample.
Although the NRO and IRAM samples show a feature of dominant warm dust,
the IRAM sample, which consists of mostly mergers or interacting galaxies,
shows evidence of very active star-formation,
whereas the NRO sample shows moderate star-formation.
Thus properties of the NRO sample are distributed 
between the IRAM sample and the other samples.


\vspace{1pc}\par
We acknowledge the works of CO-line observations
for galaxies at intermediate redshift
by Solomon et al., Sanders et al., Mirabel et al.
and Lavezzi \& Dickey. 
This research also acknowledge the NASA/IPAC Extragalactic Databese
to make use of the IRAS data.
The authors $YT$ and $MH$ acknowledge the financial support
by the Research Fellowships of the Japan Society for 
the Promotion of Science for Young Scientists.


\section*{Appendix 1.\ IRAS color-color diagram}

We discussed the FIR luminosity,
dust mass, dust temperature, and gas-to-dust ratio 
estimated from the IRAS  fluxes at 60$\mu\rm m$ and 100$\mu\rm m$
on the assumption of thermal dust emission with an emissivity
law of $n=1$.
In order to test this assumption, 
we plot a color-color diagram at IRAS  wavelengths,
of 12$\mu\rm m$, 25$\mu\rm m$, 60$\mu\rm m$ and 100$\mu\rm m$.
Figure 10a shows IRAS  color-color diagrams
of $\log (S_{12}/S_{25})$ against $\log (S_{60}/S_{100})$.
The top left panel is an IRAS  color-color diagram for all the samples.
Sauvage \& Thuan (1994) examined IRAS  color-color diagrams 
for normal {\it and} non-IR luminous galaxies, 
and discussed the dependency of the morphology of galaxies
on a color-color diagram.
The dot-and-broken quadrangle box in the top left panel of
of figure 10a represents the distribution of 
the normal and non-IR luminous galaxies with the morphology of 
Sdm - Im given by Sauvage \& Thuan (1994).
Rowan-Robinson \& Crawford (1989)
also examined IRAS  color-color diagrams
for starburst, Seyfert and quiescent spiral galaxies.
The distributions of these galaxies are shown 
in figures 10a--c 
as the solid box (quiescent spirals),
the broken box (starburst galaxies) 
and the dotted box (Seyfert galaxies), respectively.
Galaxies whose  IRAS  fluxes are known at all IRAS  bands are 
denoted by filled symbols.
Galaxies whose 12$\mu\rm m$ flux is given by an upper limit
are denoted by open symbols. 
Therefore, the real data point of the open symbols
will be put leftward.
Parts of the regions of the quiescent, 
starburst and Seyfert galaxies overlap.
Comparing each sample with the results of Sauvage \& Thuan (1994) 
and Rowan-Robinson \& Crawford (1989) in figure 10a,
it is found that the IRAM sample  and SEST sample 
data points fall near the region of Seyfert and starburst galaxies. 
The NRAO-S sample  is distributed within  
the region of quiescent spiral galaxies,
whereas some galaxies in the samples are plotted at the 
Seyfert and starburst regions.
The NRAO-L sample  is distributed in the region of 
quiescent spiral galaxies.
Although in the NRO sample  there are only two galaxies 
whose IRAS  fluxes are fixed, 
the real data point of the galaxies 
denoted by open squares will be plotted leftward.
Therefore, the NRO sample  has characteristics similar 
to quiescent spiral galaxies,
so far as the color-color diagram is concerned.
Figures 10b and c are  IRAS 
color-color diagrams for colors of $\log (S_{12}/S_{25})$,
$\log (S_{25}/S_{60})$ and $\log (S_{60}/S_{100})$.
Comparing between them, the same trends 
as figure 10a are found.

\section*{Appendix 2.\ Galactic active phenomena}

Recently it has been recognized that the majority of galaxies, 
even normal galaxies, have a supermassive black hole 
including the Milky Way 
(e.g. van der Marel 1998, and references therein)
which is suggestive that they have experienced 
galactic active phenomena 
such as starburst, active galactic nuclei (AGN) or powerful radio 
emission (e.g. Ho et al. 1995) due to supermassive black holes
at the galactic center to galaxy interactions. 
Seyfert and starburst galaxies, whose CO line has been detected,
have been discussed in terms of CO and FIR luminosities 
(e.g. Mazzarella et al. 1993; Sanders \& Mirabel 1996; 
Rigopoulou et al. 1997).
A correlation between CO and FIR luminosities
for Seyfert and starburst galaxies 
shows the same trend as for LIGs 
and is indistinguishable from them
(Heckman et al. 1989; Rigopoulou et al. 1997),
and also type 1 and type 2 Seyferts show no difference.
This suggests that FIR emission in Seyfert and starburst galaxies
is responsible for the similar origin 
({\it i.e.} dust re-radiation of starlight)
of ULIGs and normal galaxies.  
Furthermore a comparison of the luminosity ratio of FIR to CO, 
$L_{\rm FIR}/{L}_{\small{\rm CO}}'$, with the dust temperature, $T_{d}$, also shows 
the same trend as ULIGs and normal galaxies,
suggesting that the FIR emission is thermal in origin.
Mazzarella et al. (1993) examined the CO and FIR properties 
for powerful radio galaxies whose CO emission had been detected,
and also found the same trend between CO and FIR luminosities 
as ULIGs and normal galaxies,
whereas non-thermal radio power showed a large excess
compared to those objects. 
It suggests that the origin of powerful radio galaxies
may be closely related to the genesis of dust-enshrouded
quasar and classical UV-excess quasars through merging
of gas rich disk galaxies (e.g. Sanders et al. 1988a; 1988b).
This trend is also clearly found in an examination by
Sanders \& Mirabel (1996).

In order to discuss the galactic activity for the NRO sample , 
we compared CO and FIR properties of nearby
Seyfert, starburst and powerful radio galaxies 
with the NRO sample  and the ULIGs in figures 11 and 12.
The nearby CO data references from the literature are as follows:
Seyfert galaxies (Heckman et al. 1989), 
starburst galaxies (Jackson et al. 1989)
and powerful radio galaxies which have been detected in the CO-line
(Mazzarella et al. 1993).
Figure 12 shows the CO and FIR luminosities for the nearby active 
galaxies compared to the galaxies at intermediate redshift.
Figure 12 shows that 
the excess of $L_{\rm FIR}/{L}_{\small{\rm CO}}'$ for the NRO sample  is not 
directly related to the galactic activity. 
The active galaxies are  indistinguishable from the galaxies 
at intermediate redshift.



\clearpage
\section*{References}

\re
	Allen C.W.\ 1973, in Astrophysical Quantities (3rd ed.), 
	The Athlone Press
\re
	Bohlin R.C., Savage B.D., Drake J.F.\ 1978, ApJ 224, 132
\re
	Condon J.J.\ 1992, ARA\&A 30, 575
\re
	de Jong T., Brink K.\ 1987, sfig conf, 323
\re
	Devereux N.A., Young J.S.\ 1990, ApJ 359, 42
\re
	Downes D., Solomon P.M., Radford S.J.E.\ 1993, ApJ 414, L13
\re
	Elfhag T., Booth R.S., Hoeglund B., Johansson L.E.B., 
	Sandqvist A.\ 1996, A\&AS 115, 439
\re
	Fisher K.B., Huchra J.P., 
	Strauss M.A., Davis M., et al.\ 1995, ApJS 100, 69
\re
	Heckman T.M., Blitz L., 
	Wilson A.S., Armus L., Miley G.K.\ 1989, ApJ 342, 735
\re
	Hildebrand R.H.\ 1983, Quart.J.R.A.S. 24, 267
\re
	Hunt L.K.\ 1991, ApJ 370, 511
\re
	Ho L.C., Filippenko A.V.,  
	Sargent W.L.\ 1995, ApJS 98, 477
\re
	Jackson J.M., Snell R.L., 
	Ho P.T.P., Barrett A.H.\ 1989, ApJ 337, 680
\re
	Kim D.C., Sanders D.B.\ 
	1998, ApJS 119, 41
\re
	Lavezzi T.E., 
	Dickey J.M.\ 1998, AJ 116, 2672 [NRAO-L sample ]
\re
	Lonsdale C.J., 
	Helou G., Good J.C., Rice W.\ 1985, 
	in Cataloged Galaxies and Quasars
	Observed in the IRAS  Survey (Pasadena, JPL)
\re
	Mazzarella J.M., 
	Graham J.R., Sanders D.B., Djorgovski S.\ 
	1993, ApJ 409, 170
\re
	Mirabel I.F., Booth R.S., 
	Johansson L.E.B., Garay G., Sanders D.B.\ 
	1990, A\&A 236, 327 [SEST sample ]
\re
	Mooney T.J.,  
	 Solomon P.M.\ 1988, ApJ 334, L51
\re
	Rickard L.J.,   
	Harvey P.M.\ 1984, AJ 89, 1520
\re
	Rigopoulou D., 
	Papadakis I., Lawrence A., Ward M.\ 1997, A\&A 327, 493
\re
	Rowan-Robinson M., Crawford J.\ 1989, MNRAS 238, 523
\re
	Sage L.J.\ 1993, A\&A 272, 123
\re
	Sanders D.B., 
	Soifer B.T., Elias J.H., Madore B.F., Matthews K., 
	Neugebauer G., et al.\ 1988, ApJ 325, 74
\re
	Sanders D.B., 
	Soifer B.T., Elias J.H., Neugebauer G., Matthews K.\ 
	1988, ApJ 328, L35
\re
	Sanders D.B., 
	Scoville N.Z., Soifer B.T.\ 1991, ApJ 370, 158 [NRAO-S sample ]
\re
	Sanders D.B.,  
	Mirabel I.F.\ 1985, ApJ 298, L31
\re
	Sanders D.B.,  
	Mirabel I.F.\ 1996, ARA\&A 34, 749
\re
	Sauvage M., 
	Thuan T.X.\ 1994, ApJ 429, 153
\re
	Soifer B.T., Rowan-Robinson M., 
	Houck J.R., de Jong T., Neugebauer G., Aumann H.H., 
	et al.\ 1984, ApJ 278, L71
\re
	Solomon P.M., Rivolo A.R.,
	Mooney T.J., Barrett J.W., Sage L.J.\ 1986,
	in Proc. of Conference, Star Formation in Galaxies, 
	Pasadena 1986, ed C. Lonsdale, p.37
\re
	Solomon P.M.,  
	Sage L.J.\ 1988, ApJ 334, 613
\re
	Solomon P.M.,  
	Rivolo A.R.\ 1989, ApJ 339, 919
\re
	Solomon P.M., Barrett J.W.\ 1991, 
	in Dynamics of Galaxies and Their Molecular Clouds
	Distributions, 
	ed F.Combes \& F.Casoli, (Dordrecht: Kluwer), p235
\re
	Solomon P.M., Radford S.J.E., Downes D.\ 1992, 
	Nature 356, 318
\re
	Solomon P.M., Downes D., Radford S.J.E.,  
	Barrett J.W.\ 1997, ApJ 478, 144 [IRAM sample ]
\re
	Spitzer L.\  1978,
	in Physical processes in the interstellar medium,
	New York Wiley-Interscience
\re
	Stark A.A., Knapp G.R., 
	Bally J., Wilson R.W., Penzias A.A., Rowe H.E.\ 
	1986, ApJ 310, 660
\re
	Strauss M.A., Huchra J.P., 
	Davis M., Yahil A., et al.\ 1992, ApJS 83, 29 
\re
	Strong A.W., Bloemen J.B.G.M., Dame T.M., et al.\ 
	1988, A\&A 207, 1
\re
	Tinney C.G., Scoville N.Z., 
	Sanders D.B., Soifer B.T.\ 1990, ApJ 362, 473
\re
	Tutui Y., Sofue Y.\ 
	1997, A\&A 326, 915
\re
	van der Marel R.P.\ 1998,
	in Galaxy Interaction at Low and High Redshift, 
	ed D.B.Sanders, J.Barnes (Dordrecht: Kluwer)
\re
	Young J.S., Kenney J., 
	Lord S.D., Schloerb F.P.\ 1984, ApJ 287, L65
\re
	Young J.S., Schloerb F.P., 
	Kenney J.D., Lord S.D.\ 1986, ApJ 304, 443
\re
	Young J.S., Xie S., 
	Kenney J.D.P., Rice W.L.\  1989, ApJS 70, 699
\re
	Young J.S., Xie S.,  
	Tacconi L., Knezek P., Viscuso P., 
	Tacconi-Garman L., et al.\ 1995, ApJS 98, 219

\clearpage
{\bf Figure Caption}

Figure 1. --- CO-line profiles for all galaxies whose CO linewidth 
was obtained using the NRO 45-m telescope. 
The scale of intensity is the antenna temperature ($T_{\rm A}^{\ast}$).
The center of abscissa is the heliocentric radial velocity/frequency
given by optical redshift determinations.

Figure 2. --- Redshift ($cz$) and FIR luminosity ($L_{\rm FIR}$) distributions
for all the samples.
IRAM sample ({\it asterisk}), 
NRAO-S sample ({\it plus}), 
SEST sample ({\it open square}),
NRAO-L sample ({\it open circle}) and 
NRO sample ({\it filled square}).
The broken line represents an iso-flux line for a source.

Figure 3. --- Redshift ($cz$) and CO luminosity (${L}_{\small{\rm CO}}'$) distributions
for all the samples.
Symbols are the same as Fig. 2.
The symbol in parentheses denotes marginal detection,
and the symbol with a arrow denotes upper/lower limit.
The broken line represents an iso-flux line of a source.

Figure 4. --- Diagram of CO and FIR luminosities.
Symbols are the same as Fig.2.
Solid thick line indicates the average of 65 nearby spiral galaxies 
as a distance-limited sample presented by Sage (1993),
and the solid thin line indicates the average of 93 nearby spiral 
galaxies as a flux-limited sample presented by Solomon et al.(1988), 
and the dispersion is about order of 1.
The symbol in parentheses denotes marginal detection,
and the symbol with a arrow denotes upper/lower limit.
The dotted lines represent the range of $L_{\rm FIR}/{L}_{\small{\rm CO}}'$
seen in giant molecular clouds in the Milky Way
($5 \leq L_{\rm FIR}/{L}_{\small{\rm CO}}' \leq 50$). 

Figure 5. --- Diagrams of CO and FIR luminosities for each sample.
Dotted lines represent the ratio of $L_{\rm FIR}/{L}_{\small{\rm CO}}'$
of 5 and 50.

Figure 6. --- A diagram of molecular gas mass derived from CO luminosity
and dust mass derived from FIR luminosity.
Solid line is an observed gas-to-dust ratio ($\sim 570$)
for nearby spiral galaxies using IRAS flux at 100 $\mu\rm m$.
Broken line represents a typical gas-to-dust ratio 
of the Galaxy ($\sim 100$) from observations of nearby stars.
The symbol in parentheses denotes marginal detection,
and the symbol with a arrow denotes upper/lower limit.

Figure 7. --- Diagram of dust temperature and a ratio of 
FIR-to-CO luminosity.
The unit of FIR and CO luminosity are ${L}_{\odot}$ and $~{\rm K~km~s{}^{-1}~pc{}^{2}}$, 
respectively.
Solid and broken line are fitted by a power law 
for all the samples and for the data without NRO sample and IRAM sample,
respectively.
The symbol in parentheses denotes marginal detection,
and the symbol with a arrow denotes upper/lower limit.

Figure 8. --- Residual of IRAS flux ratio of 60$\mu\rm m$ $to$ 100$\mu\rm m$
($S~'_{60}/S~'_{100}$) to which $K$-correction is applied, 
{\it to} observed flux ratio ($S_{60}/S_{100}$).
The ordinate is $\log (S~'_{60}/S~'_{100}) - \log (S_{60}/S_{100})$.
Here we assumed the thermal spectral energy distribution and 
the emissivity law of $S_{\nu} \propto \lambda^{-1}$.
The lines represent different cases of the dust temperature. 

Figure 9. --- Ratio of dust mass to which $K$-correction is 
applied ($M_{\rm dust}'$) {\it to} 
that without $K$-correction ($M_{\rm dust}$)
along redshift $cz$,
where the thermal spectral energy distribution and 
the emissivity law of $S_{\nu} \propto \lambda^{-1}$ are assumed.
The lines represent different cases of the dust temperature. 

Figure 10.(a) --- IRAS color-color diagrams of the flux densities
at 12 $\mu\rm m$ $to$ 25 $\mu\rm m$
against 60 $\mu\rm m$ $to$ 100 $\mu\rm m$.
All the samples are shown in the top left panel, which is overlaid with
IRAM sample ($middle left panel$), 
NRAO-S sample ($bottom left$),
SEST sample ($top right$), 
NRAO-L sample ($middle right$) 
and NRO sample ($bottom right$).
Filled symbols represent fixed values
at all IRAS bands.
Open symbols represent that the 12 $\mu\rm m$ flux is 
given as an upper limit,
therefore, the real values should be put leftward.
The dot-and-broken quadrangle box represents the distribution of 
the normal $and$ non-IR luminous galaxies with the morphology of 
Sdm - Im given by Sauvage \& Thuan (1994).
The distributions of quiescent spiral galaxies, starburst 
and Seyfert galaxies given 
by Rowan-Robinson \& Crawford (1989) are represented
by the solid box, broken box and dotted box,
respectively.

Figure 10.(b) --- IRAS color-color diagrams of 12 $\mu\rm m$ $to$ 
25 $\mu\rm m$ 
against 25 $\mu\rm m$ $to$ 60 $\mu\rm m$.
The solid box, broken box and dotted box  represent 
the distribution of quiescent spiral galaxies, starburst galaxies
and Seyfert galaxies, respectively.

Figure 10.(c) --- IRAS color-color diagrams of 
25 $\mu\rm m$ $to$ 60 $\mu\rm m$ 
against 60 $\mu\rm m$ $to$ 100 $\mu\rm m$.
Boxes represent the same as the previous figures.  

Figure 11. --- Comparison of Seyfert, starburst and powerful radio galaxies
with NRO sample and the other samples plotted 
in Fig.4
in terms of the CO and FIR luminosities.
Type 1 Seyfert ({\it filled triangles}) 
and type 2 Seyfert ({\it open triangles}) 
are referred from Heckman et al.(1989), 
starburst galaxies ({\it filled diamonds})
are referred from Jackson et al.(1989)
and powerful radio galaxies ({\it open diamonds})
are referred from Mazzarella et al.(1993).
NRO sample and the other samples are denoted by 
crossed and small dots, respectively.

Figure 12. --- Comparison of Seyfert, starburst and powerful radio galaxies
with NRO sample and the other samples plotted 
in Fig.7
in terms of the CO-to-FIR luminosity ratio ($L_{\rm FIR}/{L}_{\small{\rm CO}}'$)
and the dust temperature ($T_{d}$).
Symbols and the references are the same 
as Fig.11.
The solid line are the regression line fitted by a power law 
for all the samples used in Fig.7.

\end{document}